----------
X-Sun-Data-Type: default
X-Sun-Data-Description: default
X-Sun-Data-Name: qgroups1
X-Sun-Charset: us-ascii
X-Sun-Content-Lines: 527

\magnification=1200
\settabs 18 \columns

\baselineskip=12 pt
\topinsert \vskip 1.25 in
\endinsert

\def\sqr#1#2{{\vcenter{\vbox{\hrule height.#2pt
 \hbox{\vrule width.#2pt height#1pt \kern#1pt
 \vrule width.#2pt} \hrule height.#2pt}}}}

\def\operp{\hbox{${\kern+.25em{\bigcirc}
\kern-.85em\bot\kern+.85em\kern-.25em}$}}

\def\lsim{\;\raise0.3ex\hbox{$<$\kern-0.75em\raise-1.1ex\hbox{$\sim$}}\;}
\def\gsim{\;\raise0.3ex\hbox{$>$\kern-0.75em\raise-1.1ex\hbox{$\sim$}}\;}
\def\no{\noindent}

\def\ce{\centerline}
\def\ve{\vfill\eject}
\def\rdots{\mathinner{\mkern1mu\raise1pt\vbox{\kern7pt\hbox{.}}\mkern2mu
 \raise4pt\hbox{.}\mkern2mu\raise7pt\hbox{.}\mkern1mu}}

\def\e e{$e^+ e^-$ }



\rightline{UCLA/99/TEP/47}
\rightline{February 2000}
\vskip1.0cm

\ce{\bf QUANTUM GROUPS AND FIELD THEORY}
\vskip.5cm

\ce{R. J. Finkelstein}
\vskip.3cm
\ce{\it Department of Physics and Astronomy}
\ce{\it University of California, Los Angeles, CA 90095-1547}
\vskip1.0cm

\no{\bf Abstract.}  When the symmetry of a physical theory describing a finite
system is deformed by replacing its Lie group by the corresponding quantum
group, the operators and state function will lie in a new
algebra describing new degrees of freedom.  If the symmetry of a field
theory is deformed in this way, the enlarged state space will again
describe additional degrees of freedom, and the energy levels will acquire
fine structure.  The massive particles will have a stringlike spectrum 
lifting the degeneracy of the point-particle theory, and
the resulting theory will have a non-local description.  Theories of this
kind naturally contain two sectors with one sector lying close to the standard
theory while the second sector describes particles that should be more
difficult to observe.
\vskip1.0cm

\no PACS numbers: 81R50, 81T13

\ve

\line{{\bf 1.  Introduction.} \hfil}
\vskip.3cm

Since the Lie groups may be considered as degenerate forms of the
quantum groups,$^1$ it may be of interest to
generalize the symmetry of a physical
theory by replacing its Lie group by the corresponding quantum
group.$^2$  When this generalization is attempted for the quantum
mechanics of finite systems, such as the harmonic oscillator or the hydrogen atom, it is found that the state space must be expanded to describe additional degrees of freedom in a manner rather similar to the way that spin fine structure is added to an atomic structure.$^3$ 
When the corresponding generalization of field theory is carried out, the expanded state space may be interpreted to describe additional degrees of freedom associated with the spatial extension of the field quanta, or elementary particles.$^3$
While the fine structure of stringlike theories results from explicit geometrical postulates about non-local structure, the generalization of gauge theories to quantum groups leads to fine structure without the necessity of supplementary geometric input.
If successful this procedure would resemble the manner in which the algebraic formulation of Pauli spin replaces the geometric
picture of a spinning electron.  
\vskip.5cm

\line{{\bf 2. The Quantum Group $SL_q(2)$.} \hfil}
\vskip.3cm

We base our work on the simplest non-trivial example: $SL_q(2)$.
Its 2-dimensional representation, $T$, may be defined as
follows:
$$
\eqalignno{&T\epsilon T^t = T^t\epsilon T = \epsilon & (2.1) \cr
&{\rm det}_q T = 1 & (2.2) \cr}
$$
\no where $\epsilon$ is a 2-dimensional representation of the
imaginary unit:
$$
\epsilon = \left(\matrix{0 & q_1^{1/2} \cr
-q^{1/2} & 0 \cr} \right) \quad
q_1 = q^{-1}~. \eqno(2.3)
$$
\no Here $T^t$ is the transposed matrix and det$_qT$ is defined
by
$$
\epsilon_{ij}T_{im}T_{jn} = \epsilon_{mn}{\rm det}_qT~. \eqno(2.4)
$$
\no Set
$$
T = \left(\matrix{\alpha & \beta \cr
\gamma & \delta \cr} \right)~. \eqno(2.5)
$$
\no Then by (2.1) and (2.2)
$$
\eqalign{\alpha\beta &= q\beta\alpha \cr
\delta\beta &= q_1\beta\delta \cr} \quad
\eqalign{\alpha\gamma &= q\gamma\alpha \cr
\delta\gamma &= q_1\gamma\delta \cr} \quad
\eqalign{\alpha\delta &- q\beta\gamma = 1 \cr
\delta\alpha &- q_1\beta\gamma = 1 \cr} \quad
\eqalign{\beta\gamma &= \gamma\beta \cr
\hfil \cr} \eqno(2.6)
$$
\no Consider now a matrix realization of the algebra (2.6) and
set
$$
\delta = \bar\alpha~, \quad  
\beta = \bar\beta \quad \hbox{and} \quad
\gamma = \bar\gamma  \eqno(2.7)
$$
\no where the bar signifies Hermitian conjugation.

Then
$$
\eqalign{\alpha\beta &= q\beta\alpha \cr
\bar\alpha\beta &= q_1\beta\bar\alpha \cr} \quad
\eqalign{\alpha\gamma &= q\gamma\alpha \cr
\bar\alpha\gamma &= q_1\gamma\bar\alpha \cr} \quad
\eqalign{\alpha\bar\alpha &- q\beta\gamma = 1 \cr
\bar\alpha\alpha &- q_1\beta\gamma = 1 \cr} \quad
\eqalign{\beta\gamma &= \gamma\beta \cr
\hfil \cr} \eqno(2.8)
$$
\no Under Hermitian conjugation, Eqs. (2.8)
imply
$$
q = \bar q \eqno(2.9)
$$
\no so that $q$ is a real number.  Since $\beta$ and $\gamma$
commute, they have a common set of eigenstates $|nm\rangle$.
Let the ground states of $\beta$ and $\gamma$ be $|om\rangle$
and $|no\rangle$ respectively, where
$$
\eqalign{\alpha|om\rangle &= \alpha|no\rangle = 0 \cr
\beta|om\rangle &= b|om\rangle \cr
\gamma|no\rangle &= c|no\rangle~. \cr}
\eqno(2.10)
$$
\no Define the state $|nm\rangle$ by the recursive
relations:
$$
\eqalignno{\bar\alpha|nm\rangle &= 
\lambda_{nm}|n+1,m+1\rangle & (2.11) \cr
\alpha|nm\rangle &= \mu_{nm}|n-1,m-1\rangle~. & (2.12) \cr}
$$

By (2.8) and (2.10)
$$
\eqalignno{&\beta|nm\rangle = q^nb|nm\rangle & (2.13) \cr
&\gamma|nm\rangle = q^mc|nm\rangle & (2.14) \cr
&bc = -q & (2.15) \cr
&|\lambda_{nm}|^2 = 1 - q^{n+m+2} & (2.16) \cr
&|\mu_{nm}|^2 = 1-q^{n+m} & (2.17) \cr}
$$
\no By the preceding equations
$$
q<1~. \eqno(2.18)
$$

\vskip.5cm

\line{{\bf 3. Mass Spectrum in a Global Theory.} \hfil}
\vskip.3cm

Suppose that there is a mass term of the following form:
$$
M\tilde\psi\epsilon\psi \eqno(3.1)
$$
\no where $\psi$ is a fundamental representation of the $T$-group:
$$
\eqalignno{\psi(x)^\prime &= T\psi(x) & (3.2) \cr
\tilde\psi(x)^\prime &= \tilde\psi(x) T^t & (3.3) \cr}
$$
\no Then by (2.1) the mass term is invariant.  Here $T$ is
position independent.

The new idea that we are exploring is that the fields and
the states lie in the $T$-algebra.  Therefore suppose that
$\psi$ and $\tilde\psi$ have the following expansions
$$
\eqalignno{\psi &= \psi^{(\beta)}\beta + \psi^{(\gamma)}\gamma
& (3.4) \cr
\tilde\psi &= \beta\tilde\psi^{(\beta)} + \gamma\tilde\psi^{(\gamma)} & (3.5) \cr}
$$
\no where $\psi^{(\beta)}$ and $\psi^{(\gamma)}$ as well as
$\tilde\psi^{(\beta)}$ and $\tilde\psi^{(\gamma)}$ are two-dimensional vectors
that do not lie in the algebra, and are orthogonal:
$\tilde\psi^{(\beta)}\epsilon\psi^{(\gamma)} = \tilde\psi^{(\gamma)}\epsilon
\psi^{(\beta)} = 0$.
Note that $\tilde\psi$ is not $\psi^t$.  Note that the action of $T$ on $\psi$ in (3.4)
will carry $\psi$ into other parts of the algebra (containing an infinite
number of terms of the form $\beta^n\gamma^m\alpha^s\bar\alpha^\ell$).  The
general $\psi$ therefore contains an infinite number of modes and Eq. (3.4)
can be posited only in a special gauge. 
The expectation value of (3.1) in the state $(nm)$ is
$$
M\bigl[(\tilde\psi^{(\beta)}\epsilon\psi^{(\beta)})
\langle nm|\beta^2|nm\rangle + (\tilde\psi^{(\gamma)}\epsilon
\psi^{(\gamma)})\langle nm|\gamma^2|nm\rangle\bigr]~.
\eqno(3.7)
$$
\no Then the contribution of this term to the spectrum is, by (2.13) and (2.14)
$$
M[f^{(\beta)}q^{2n}b^2 + f^{(\gamma)}q^{2m}c^2] \eqno(3.8)
$$
\no where
$$
f^{(\beta)} = \tilde\psi^{(\beta)}\epsilon\psi^{(\beta)}
\eqno(3.9)
$$
\no and there is a corresponding expression for $f^{(\gamma)}$.
Set
$$
f^{(\beta)} b^2 = {1\over V}~. \eqno(3.10)
$$
\no Then (3.8) may be written by (2.15) as follows:
$$
M\biggl[q^{2n}{1\over V} + q^{2m+2}(f^{(\beta)}f^{(\gamma)})V\biggr]~.
\eqno(3.11)
$$

The spectrum associated with $M$ is inverted since $q<1$.  Eq. (3.11)
bears some resemblance to the spectrum of the toroidally compactified
string with its related large small ($T$) duality.  By (2.11) however
$n=m$.  To obtain states $|nm\rangle$ with $n\not= m$ one may go to a higher
rank quantum group.

The new states and levels may be likened to new states of polarization
that depend on the tensor nature of $\psi$.

While $q$ is dimensionless, there is no restriction on the dimensionality
of the eigenvalues $b$ and $c$.  If $b$ is a length, and $V$ is a volume, then
by (3.10) dim $f^{(\beta)} = L^{-5}$ and dim $f^{(\gamma)} = L^{-1}$.

For given $n$, the mass is minimized for the characteristic volume
$$
\bar V = (q^2f^{(\beta)}f^{(\gamma)})^{-1/2}~. \eqno(3.12)
$$
\no Since $n=m$, (3.12) corresponds to the self-dual solution of (3.11).

In the limit of a simple field theory this spectrum collapses to
a single level and the field quanta are point particles.
In the present case the existence of any kind of mass spectrum 
may be interpreted to imply extension in configuration space.
Assuming that the mass of a field quantum is given by (3.11) 
one may gain a qualitative idea of its spatial extension by
noting that $q^n$ differs little from a 
$\langle n\rangle_q$ spectrum---essentially the spectrum of a
$q$-isotropic oscillator, with zero angular momentum.  The wave
functions of a $q$-oscillator are $q$-Hermite functions.  The natural scale
of the soliton would be fixed by (3.14).

If $\psi$ is a Dirac spinor then the usual Lorentz invariant
term is
$$
M\psi^tC\psi \eqno(3.15)
$$
\no where $C$ is the charge conjugation matrix that intertwines
$L^t$ and $L^{-1}$
$$
L^tCL = LCL^t = C~. \eqno(3.16)
$$

If it is supposed that the Lorentz symmetry is broken, one
possibility is that $L$ is replaced by the $q$-Lorentz group
$(L_q)$ which is defined by its spin representation as follows:
$$
\epsilon~{\rm det}_q L_q = L_q^t\epsilon L_q \qquad
{\rm det}_q L_q = 1 \eqno(3.17)
$$
\no Then (3.15) is replaced by
$$
M\tilde\psi\epsilon\psi \eqno(3.18)
$$
\no and the new charge conjugation matrix is $\epsilon$
itself.

If this term is to be invariant under independent Lorentz and
$T$ transformation, we may write
$$
M\tilde\psi C\epsilon\psi \eqno(3.19)
$$
\no where $\tilde\psi^\prime = \tilde\psi L^t$ under Lorentz
transformations and $\tilde\psi^\prime = \tilde\psi T^t$ under
$T$ transformations.
\vskip.5cm

\line{{\bf 4. Gauge Theory.} \hfil}
\vskip.3cm

Much of the standard group theory, including orthogonality of
irreducible representations, Clebsch-Gordan rules, etc., may be
carried over to quantum groups.  Closure, however, requires
$$
(T_1T_2)^t = T_2^tT_1^t \eqno(4.1)
$$
\no but (4.1) is guaranteed only if the matrix elements of 
$T_1$ commute with those of $T_2$.  To ensure this property one
may take $T_1$ and $T_2$ at spatially (causally) separated
points with respect to the light cone.  The resulting groupoid
is non-local.

When $T(x)$ depends on position, we may adopt the following Lagrangian which is both Lorentz and $T$-invariant
$$
L = -{1\over 4} \sum_\alpha L(\alpha)\vec F_{\mu\nu}\vec F^{\mu\nu}R(\alpha) +
i~\tilde\psi C\epsilon\gamma^\mu\vec\nabla_\mu\psi +
{1\over 2} 
\bigl[(\tilde\varphi\buildrel \leftarrow\over {\nabla}_\mu)
\epsilon(\vec\nabla_\mu\varphi) + \tilde\varphi\epsilon
\varphi\bigr] \eqno(4.2)
$$
\no where
$$
\eqalign{L(\alpha)^\prime &= L(\alpha)T^{-1} \cr
R(\alpha)^\prime &= TR(\alpha) \cr
(L\epsilon)^\prime &= (L\epsilon)T^t \cr
\hfil \cr}
\eqalign{\varphi^\prime &= T\varphi \cr
\tilde\varphi^\prime &= \tilde\varphi T^t \cr
\psi^\prime &= T\psi \cr
\tilde\psi^\prime &= \tilde\psi T^t \cr} \quad
\eqalign{(\vec\nabla\varphi)^\prime &= T(\vec\nabla\varphi) \cr
(\tilde\varphi \buildrel \leftarrow\over\nabla)^\prime &=
(\varphi \buildrel\leftarrow\over\nabla)T^t \cr
(\vec\nabla\psi)^\prime &= T(\vec\nabla\psi) \cr
(\tilde\psi \buildrel\leftarrow\over\nabla)^\prime &=
(\tilde\psi\buildrel\leftarrow\over\nabla)T^t \cr}
\eqno(4.3)
$$
$$
\eqalign{\vec\nabla_\mu^\prime &= T\vec\nabla_\mu T^{-1}~, \cr
\buildrel\leftarrow\over{\nabla}^\prime_\mu &=
(T^t)^{-1}\buildrel\leftarrow\over{\nabla}_\mu T^t~, \cr} \quad
\eqalign{
\vec\nabla_\mu &= \vec\partial_\mu + \vec A_\mu~, \cr
\buildrel\leftarrow\over{\nabla}_\mu &= 
\buildrel\leftarrow\over{\partial}_\mu + \buildrel\leftarrow\over{A}_\mu~, \cr}
\quad
\eqalign{\vec F_{\mu\nu} &= (\vec\nabla_\mu ,\vec\nabla_\nu) \cr
\buildrel\leftarrow\over{F}_{\mu\nu} &= (\buildrel\leftarrow\over{\nabla}_\mu ,
\buildrel\leftarrow\over{\nabla}_\nu) \cr}
$$
\no Kinetic terms in $L(\alpha)$ and $R(\alpha)$, as well as other possible terms in
$\buildrel \leftarrow\over A$ have not been expressed in (4.2).

The invariance of the Lagrangian requires distinct left and
right fields because $F$ does not commute with $T$.  In the limit $q=1$
of (4.2) the $L$ and $R$ fields may be summed out as follows:
$$
\sum_\alpha L_i(\alpha)R_j(\alpha) = \delta_{ij} \eqno(4.4)
$$
\no where the sum is over the complete set of left and right fields.  Then
$$
\lim_{q\to 1} \sum_{\alpha} L_i(\alpha) (FF)_{ij} R_j(\alpha) =
{\rm Tr}~FF~. \eqno(4.5)
$$

In this limit (4.2) becomes the usual Yang-Mills Lagrangian for $SU(2)$.
In this limit also the internal structure described by the $q$-algebra of
course disappears.  The energy levels of the limiting Yang-Mills theory are
therefore highly degenerate when viewed from within the $q=1$ theory.  This
degeneracy is lifted when $q$ is turned on.

\vskip.5cm

\line{{\bf 5. Representation of the Free Fields.} \hfil}
\vskip.3cm

After expanding the space of one-particle states we may adopt the conventional
representations of the free scalar and spinor fields:
\vskip.3cm

scalar
$$
\varphi(x) = \biggl({1\over 2\pi}\biggr)^{3/2}\int {d\vec p\over (2p_o)^{1/2}}
\sum_s u(p,s)\bigl[e^{-ipx}a(p,s) + e^{ipx}\bar a(p,s)\bigr]\tau_s
\eqno(5.1)
$$

spinor
$$
\psi(x) = \biggl({1\over 2\pi}\biggr)^{3/2}
\int {d\vec p\over (2p_o)^{1/2}}\sum_{r,s}
\bigl[u(p,r,s)e^{-ipx}a(p,r,s) + v(p,r,s)e^{ipx}\bar b(p,r,s)\bigr]\tau_s
\eqno(5.2)
$$
\no where the only new element added to the conventional expansions is the sum
over $\tau_s$ where $\tau_s$ lies in the $q$-algebra.

The corresponding choice for the vector fields offers more possibilities even
if we are guided by the standard non-Abelian theory employing the following
representation in the $SU(2)$ theory for the vector $W_\mu$:
$$
W_\mu = W_\mu(+)\tau(-) + W_\mu(-)\tau(+) + W_\mu(3)\tau_3~. \eqno(5.3)
$$

In the $SU_q(2)$ theory one option is based on the following correspondence
between the $q$-algebra and the Cartan subalgebra of $SU(2)$:
$$
\bar\alpha\sim E_+ \quad \alpha\sim E_- \quad \hbox{and} \quad
\beta,\gamma\sim H \eqno(5.4)
$$
\no Then one may propose
$$
W_\mu^{(q)} = W_\mu^+\alpha + W_\mu^-\bar\alpha +
W_\mu^{(\beta)}\beta + W_\mu^{(\gamma)}\gamma \eqno(5.5)
$$
\no where the coefficients are conventional.  However, one may also deform
the Lie algebra of $SU(2)$.  Then the generators of the $q$-Lie algebra
satisfy
$$
\eqalign{(J_3^{(q)},J_\pm^{(q)}) &= \pm J_\pm^q \cr
(J_+^{(q)},J_-^{(q)}) &= {1\over 2}[2J_3^{(q)}]_q  \cr}\eqno(5.6)
$$
\no where
$$
[2J_3^{(q)}]_q = {q^{2J_3^q}-q^{-2J_3^q}\over q-q^{-1}}
$$
\no and instead of (5.5) one may choose
$$
W_\mu^{(q)} = W_\mu^{(+)}J(-)^q + W_\mu^{(-)}J(+)^q + W_\mu(3)J_3^q~. \eqno(5.7)
$$

In constructing the $q$-theory one may choose either option.  It is more
natural, however, to take the view that the two options are not mutually
exclusive and that the $q$-theory implies a deformation of both the group
($g$) and the algebra ($u$).

The $q$-theory describing deformations of both $g$ and $u$ would therefore
be composed of two sectors.  The first sector would contain the new
particles lying in the algebra of $(\alpha,\bar\alpha,\beta,\gamma)$.  The
second sector would lie close to the standard theory but would predict
differences from the standard theory that depend on $q$ and these predictions
would be accessible by perturbation theory since $q<1$.  In the $q=1$
correspondence limit one would recover the standard theory while the new
particles lying in the first sector would disappear as $q$ approaches unity
and in any case would be more difficult to detect.

I thank C. Fronsdal and V. S. Varadarajan for useful comments. 
  
\vskip.5cm

\line{{\bf References.} \hfil}
\vskip.3cm

\item{1.} N. Yu. Reshetikhin, L. A. Takhtadzhyan, and
L. D. Fadeev, Leningrad Math. J. {\bf 1} (1990) No. 1.
\item{2.} The literature on quantum groups is extensive.  See
for example, {\it Quantum Groups}, edited by T. Curtwright,
D. Fairlee and C. Zachos, World Scientific (1991).
\item{3.} R. Finkelstein, Observable Properties of $q$-Deformed
Physical Systems, Lett. Math. Phys. {\bf 49}, 105 (1999); Gauged $q$-Fields,
hep-th/9906135; Spontaneous Symmetry Breaking of $q$-Gauge Field
Theory, hep-th/9908210.

\end
\bye

\end
\bye